\documentclass[
 aps, 
 pra,
 twocolumn,
 reprint,
 superscriptaddress,
 ]{revtex4-2}

\usepackage{amsmath}
\usepackage{amssymb}
\usepackage{graphicx}
\usepackage{bm}
\usepackage{color}
\usepackage{subfigure}

\usepackage{mathrsfs}
\usepackage[colorlinks, 
            linkcolor=blue, 
            anchorcolor=blue, 
            citecolor=blue,
            ]{hyperref}
  \usepackage{orcidlink}

\begin{document}

\title{High-fidelity and robust controlled-Z gates implemented with Rydberg atoms via echoing rapid adiabatic passage}

\author{Ming Xue~\orcidlink{0000-0002-6156-7305}}
\thanks{These authors contributed equally.\\\href{mailto:mxue@nuaa.edu.cn}{mxue@nuaa.edu.cn}}
\affiliation{Department of Physics, Nanjing University of Aeronautics and Astronautics, Nanjing 211106, China}
\affiliation{Key Laboratory of Aerospace Information Materials and Physics (NUAA), MIIT, Nanjing 211106, China}

\author{Shijie Xu~\orcidlink{0009-0005-8782-2457}}
\thanks{These authors contributed equally.\\\href{mailto:mxue@nuaa.edu.cn}{mxue@nuaa.edu.cn}}
\affiliation{Department of Physics, Nanjing University of Aeronautics and Astronautics, Nanjing 211106, China}
\affiliation{Key Laboratory of Aerospace Information Materials and Physics (NUAA), MIIT, Nanjing 211106, China}

\author{Xinwei Li~\orcidlink{0009-0005-9677-0011}}
\email{xinwei.li1991@outlook.com}
\affiliation{Beijing Academy of Quantum Information Sciences, Xibeiwang East Road, Beijing 100193, China}

\author{Xiangliang Li~\orcidlink{0000-0002-7251-7346}}
\email{lixl@baqis.ac.cn}
\affiliation{Beijing Academy of Quantum Information Sciences, Xibeiwang East Road, Beijing 100193, China}

\date{\today}

\begin{abstract} 
  High-fidelity and robust quantum gates are essential for quantum information processing,
  where neutral Rydberg atoms trapped in optical tweezer arrays serving as a versatile platform for the implementation.
  We propose a rapid adiabatic passage (RAP) scheme for achieving a high-fidelity 
  controlled-Z (CZ) gate on a neutral atom Rydberg platform. 
  Utilizing only global laser dressing, our scheme involves echoing two identical RAP pulses within the Rydberg blockade regime to realize a CZ gate
  and can be readily extended to a C$^k$Z gate with additional qubits.
  We predict a CZ gate with fidelity over 0.9995 using akali-atom parameters,
  and a CCZ gate with fidelity exceeding 0.999. 
  Moreover, the direct utilization of echoing RAP pulses enables the implementation of a four-bit CCCZ gate 
  at fidelity over 0.996 without further optimization.
  The proposed scheme, remarkably robust to variations in driving fields and realistic decoherence effects,
  holds promise for future quantum information processing applications.
\end{abstract}
\maketitle

\section{Introduction}
Rydberg-atom arrays, leveraging strong interactions and adaptable geometries, 
have become a leading platform for quantum computing and simulation, 
with neutral atoms trapped in optical tweezers serving as a powerful and promising implementation~\cite{Wu_2021, Shi_2022,Adams_2020, Bluvstein24suppressing}.
Integrated with coherent rearrangement methods, 
this yields a versatile and scalable architecture for quantum processing.
In a variety of experiments utilizing alkali-metal or alkaline-earth atom qubits, 
there have been successful demonstrations of high-fidelity gates for single- and two-qubit operations~\cite{levine2019parallel,McDonnel22Demonstration,Graham23midcircuit, finkelstein2024universal,cao2024multi,peper2024spectroscopy}, 
with qubits encoded in long-lived hyperfine or optical-metastable states~\cite{Li2022Coherent,Pelegri_2022,Ma2023}. 

Neutral-atom two-qubit gate fidelities lag behind those achieved in superconducting 
systems~\cite{Barends14superconducting,linke2017experimental,Setiawan23Fast,Li23time} 
and ion traps~\cite{Ballance16high,Leu23fast,Shapira23Robust}.
Although high-fidelity entangling gate protocols for neutral-atom platforms have shown significant progress, 
there remains a gap between their predicted performance and the best experimental results. 
However, the recent advancements are particularly encouraging, 
with demonstrations achieving over 0.9935~\cite{finkelstein2024universal}, 
as high as 0.994~\cite{peper2024spectroscopy} and 0.995 for CZ gate fidelity~\cite{Evered2023, Ma2023}, 
and 98.3\% raw Bell state fidelity~\cite{cao2024multi}. 
These gates are now approaching the fidelity limits set by Rydberg decay.
Achieving robust and high-fidelity two-qubit and multi-qubit gates with more than one control qubit 
remains challenging due to limited Rydberg lifetimes~\cite{Saffaman10rmp,levine2019parallel,Zhuo22High,Evered2023}.

The Rydberg blockade mechanism is key to implementing controlled multi-qubit gates on Rydberg array platforms, 
leveraging strong Rydberg-Rydberg interactions to inhibit the excitation of adjacent atoms to high-lying Rydberg states~\cite{Jaksch00fast, shen2019construction, wu2020effective}.
While adiabatic evolution is commonly employed for quantum information processing with Rydberg atoms~\cite{Beterov_2020, Keating13Adiabatic, Mitra2020Robust, Mitra23neutral}, 
its requirement for prolonged evolution times can lead to increased decoherence noise, 
potentially compromising the protocol robustness and limiting its fidelity.
Techniques like shortcuts to adiabaticity or transitionless quantum driving~\cite{Shortcut2019Gu,Li21multiple,Bosch2023Shortcut,Dalal23two} have been proposed to accelerate the process, 
but their experimental application is limited by the need for precise control of the system Hamiltonian. 
As an alternative, rapid adiabatic passage (RAP)
can be employed to facilitate the state transitions~\cite{RANGELOV20101346,Vitanov17RMP,Carrasco24Dicke}.
This method involves sweeping transition frequencies through resonance, 
similar to the approach used in the Landau-Zener problem~\cite{cohen1986quantum,Niranjan20PRA}.
Additionally, alternative gate schemes have been proposed~\cite{Muller09mesocopic,Mitra2020Robust,Keating15robust,Mohan23robust,Sola24two,Li24high,Jandura23PRXQ,Jin24Geometric,Song24fast}, and some of these have already been experimentally implemented~\cite{McDonnel22Demonstration,chow2024circuit}.
To achieve high-fidelity gate protocols and quantum state manipulation, 
pulse-shaping and quantum optimal control techniques can be employed~\cite{saffman2020symmetric, Pelegri_2022, McDonnel22Demonstration, Pagano22error,Chang_2023,Kang2022Nonadiabatic,Mao23PRA}. 
However, due to the $\sqrt{N}$ scaling of the collective Rabi frequency for Rydberg excitation within the blockade regime, 
many protocols necessitate reoptimization for an extended number of qubits~\cite{levine2019parallel, Pelegri_2022, Jandura2022timeoptimaltwothree}.

In this study, we utilize echoing RAP pulses with only global laser dressing (symmetric driving of both atoms without the need for individual addressing) on a Rydberg atom platform
to construct high-fidelity two-qubit CZ gates.
Different from the poineering works of Refs.~\cite{saffman2020symmetric,Beterov_2020,Pelegri_2022}, 
our echoing RAP pulses exhibit continuous profiles without instantaneous laser frequency changes
and all driving fields are turned off at the end of gate duration.
We employ two identical RAP pulses that guide specific 
input states through a complete cycle on the Bloch sphere, accumulating a geometric phase of $\pi$ 
while canceling out dynamical phases.
This approach not only facilitates the precise implementation of high-fidelity CZ gates 
based on realistic parameters from alkali and alkaline-earth atom experiments,
but also directly extends to C$^k$Z gates for systems up to four qubits, maintaining robust performance
even without further optimization.

The rest of the paper is structured as follows.
In Section~\ref{sec:model_protocol}, we introduce the model Hamiltonian governing Rydberg-mediated gate dynamics between two trapped atoms 
and describe our echoing protocol for realizing CZ and CCZ gates.
Section~\ref{sec:fidelity} is dedicated to assessing the fidelity and robustness of our protocol. 
We begin by evaluating its performance using parameters relevant to alkali and alkaline-earth atoms in Sec.~\ref{subsec:akali}.
Following this, in Sec.~\ref{subsec:roubust}, we investigate the impact of thermal motion and laser intensity variations on the performance of CZ and CCZ gates.
Lastly, in Sec.~\ref{sec:conclusion} we conclude with discussions.

\begin{figure}
  \centering
  \includegraphics[width=1\columnwidth]{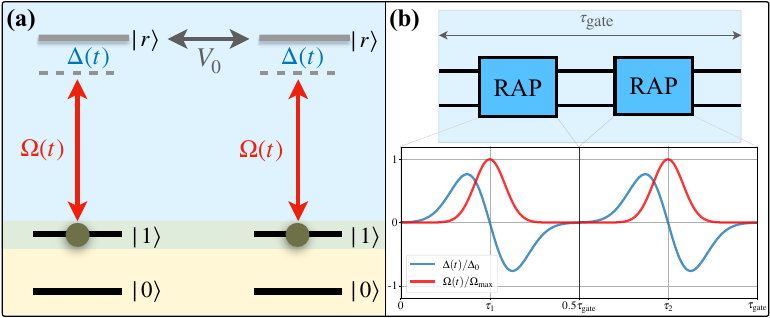}
  \caption{(a) Atomic level structure of two atom qubits featuring Rydberg-Rydberg 
  interactions between the $|r\rangle$ states. 
  (b) Schematic of a two-qubit CZ gate achieved via two echoing RAP pulses.
  The qubit is encoded in two hyperfine states, $|0\rangle$ and $|1\rangle$. 
  A global laser only dresses the ground state $|1\rangle$, 
  coupling it to the high-lying Rydberg state $|r\rangle$, 
  with Rabi frequency $\Omega$ and detuning $\Delta$.}
\label{fig:scheme}
\end{figure}

\section{Model and echoing-RAP Scheme}\label{sec:model_protocol}
\subsection{Hamiltonian and the RAP pulse}
We first consider a pair of atoms with Rydberg interaction in Fig.~\ref{fig:scheme}(a), 
each atom consists of two hyperfine ground states $|0\rangle$ and $|1\rangle$ being a computational qubit,
as well as a Rydberg excited state $|r\rangle$.
Only state $|1\rangle$ is coupled to the Rydberg state $|r\rangle$ with Rabi frequency $\Omega(t)$
and detuning $\Delta(t)$. 
These two atoms have same detunings and Rabi frequencies,
with Hamiltonian in the interaction picture~\cite{PhysRevA.107.043311,Pelegri_2022}:
\begin{equation}\label{eq:Ht}
  \hat{H}(t) = \Delta(t) \sum_{i=1}^2 |r\rangle_i \langle r|_i + \frac{\Omega(t)}{2} \sum_{i=1}^2 (|r\rangle_i \langle 1|_i+ {\rm H.c.}) + \hat{H}_{\rm I},
\end{equation}
where $\hat{H}_{\rm I} = V_0 |r\rangle_1 \langle r|_1 \otimes |r\rangle_2 \langle r|_2$,
with the Rydberg interaction strength $V_0$ depending on the distance between Rydberg atoms and the type of interaction. 
Our scheme requires only global dressing of the qubit within the blockade regime.
Both  $\Omega(t)$ and $\Delta(t)$ are time-dependent pulses, 
and their profiles will be optimized in our subsequent echoing RAP scheme to realize the target C$^k$Z gate.

Figure \ref{fig:scheme}(b) depicts a circuit representation of utilizing echoing RAP pulses to realize a 
two-qubit CZ gate, with a total gate time $\tau_{\rm gate}$.
Taking dimensionless time $\tau= \Omega_{\text{max}} t$ ($\Omega_{\text{max}}$ as the maximum Rabi frequency),
we initialize the pulse profiles for the Rabi frequency $\Omega(\tau)$ 
and detuning $\Delta(\tau)$ using ansatzes~\cite{RANGELOV20101346}: 
\begin{align} \label{eq:omega_delta}
  \Omega(\tau) &=
  \Omega_{\text{max}} e^{-[(\tau-\tau_k)/\tau_R]^2},   \\
  \Delta(\tau) &= -\Delta_0  (\tau-\tau_k) (e^{-[(\tau-\tau_k)/\tau_D]^2}-a)/(1-a), \nonumber
\end{align}  
where the interval $(k-1)/2 \leq \tau/\tau_{\rm gate} < k/2$ corresponds to the $k$-th ($k=1,2$) RAP  pulse.
Here, $\tau_{k} \equiv (2k-1)\tau_{\rm gate}/4$ denotes the central times of each RAP pulse as shown in Fig.~\ref{fig:scheme}(b).
And $a=\exp[-(\tau_\text{gate}/4\tau_D)^2]$, which is chosen to ensure that the detuning is zero at the start and end of each pulse.
The RAP pulse widths  $\{\tau_R, \tau_D\}$, detuning amplitudes $\Delta_0$ and gate time $\tau_{\rm gate}$ 
are optimized to maximize the fidelity of the engineered CZ gate,
subject to the constraint of the total gate duration $\tau\in[0,\tau_{\rm gate}]$.
The pulses in our scheme are applied uniformly and continuously across all atoms, 
potentially removing the requirement for rapid laser switching between different spatial locations. 
This could benefit gate operations in large qubit arrays and enable the simultaneous execution of multiple gates for different pairs of atoms~\cite{Schine2022Long,Shaw24multi}.

We compute the fidelity defined as:
\begin{equation}\label{eq:fidelity}
\mathcal{F} = |\langle\Psi| \mathcal{U}_{\text{C}^k\text{Z}}^\dagger \, \rho_f \, \mathcal{U}_{\text{C}^k\text{Z}} |\Psi\rangle|^2,
\end{equation}
where $\mathcal{U}_{\text{C}^k\text{Z}}$ denotes the unitary operator for the ideal C$^k$Z gate, 
and $\rho_f$ represents the final density matrix describing the actual evolution of the system under the influence of the echoing RAP pulses, 
as governed by the Hamiltonian $\hat{H}(t)$ defined in Eq.~(\ref{eq:Ht})-(\ref{eq:omega_delta}).
This fidelity could correspond to the Bell state fidelity in the two-qubit case~\cite{Evered2023}.
The state $|\Psi\rangle$ is chosen to be a symmetric superposition of all possible computational basis states, 
represented as $|\Psi\rangle = 1/\sqrt{2^n}\sum_{i,j,k,\dots=0}^{1} |ijk\dots\rangle$ for an $n$-qubit system. 
This choice of initial state makes the fidelity in Eq.~(\ref{eq:fidelity}) a sensitive indicator of both population leakage 
and relative phase errors across all computational states~\cite{Pelegri_2022}.

\subsection{C$^k$Z gate with echoing RAP pulses}
\begin{figure}
  \centering
  \includegraphics[width=1\columnwidth]{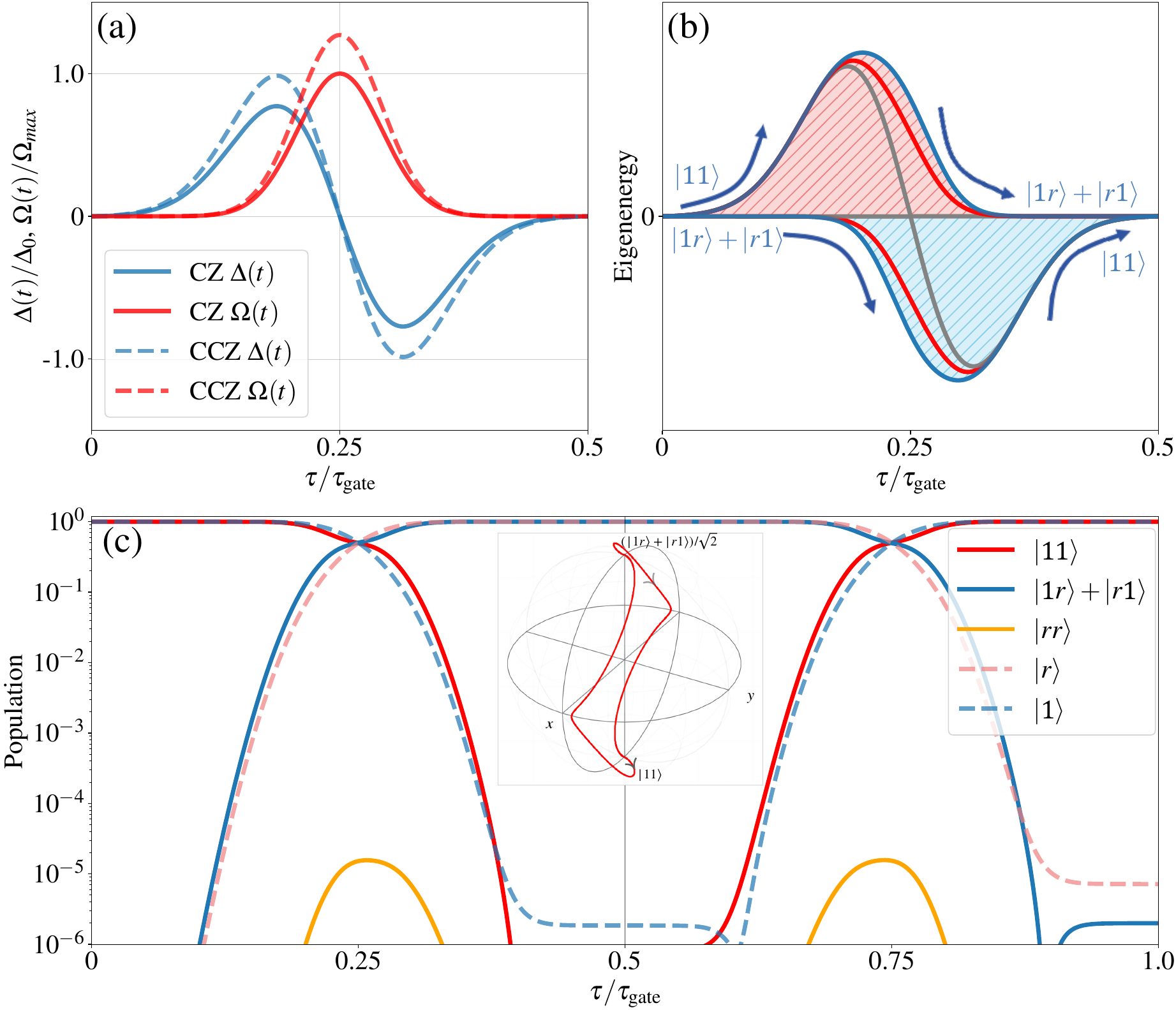}
  \caption{ CZ and CCZ gate via echoing RAP scheme. 
  (a) The profile of RAP pusle, $\Delta(t)$ and $\Omega(t)$, for CZ and CCZ gate construction.
  (b) The instantaneous eigenenergies of Hamiltonian $H(t)$ during one RAP pulse: 
  the spectra of bare states(gray lines), blue and red solid lines for the double- and single-atom eigenstates, respectively.
  The shaded area represents the dynamical phase accumulated for double-atom case. 
  (c) The atomic population evolution during the complete gate time $\tau_{\rm gate}$. 
  The solid line represents the states evolution of double-atom case, 
  and the dashed line represents the single-atom case. 
  Inset: Bloch sphere trajectory of the dynamical state.
  Parameters are $\Delta_0=0.55 \Omega_{\rm max}$, $(\tau_R, \tau_D)=(0.564,0.846)$, $\tau_{\rm gate}=\Omega_{\rm max} t=9.4$, $V_0=127\Omega_{\rm max}$ for the CZ gate, 
  and $\Delta_0=0.70 \Omega_{\rm max}$, $(\tau_R, \tau_D)=(0.72,1.08)$, $\tau_{\rm gate}=12.0$, $V_0=100\Omega_{\rm max}$ for the CCZ gate.
  }
\label{fig:CZgate}
\end{figure}

The specific ansatz utilized for the RAP pulse in Fig.~\ref{fig:CZgate}(a) facilitates coherent 
population inversion among the adiabatic eigenstates of the Hamiltonian in Eq.~(\ref{eq:Ht}).
This is illustrated in Fig.~\ref{fig:CZgate}(b), depicting the instantaneous eigenspectra of $\hat{H}(t)$ 
during one RAP pulse, which includes the double-atom (involving both atoms in the evolution, marked with solid blue) and single-atom (only one atom participates in the evolution, marked with solid red) spectra.
The significant energy gap between instantaneous eigenenergies protect the adiabatic condition to be satisfied~\cite{PhysRevApplied.18.044042},
as revealed in Fig.~\ref{fig:CZgate}(c) for input $|11\rangle$ that only negligible population leakages to other states.
States $|00\rangle$ and $(|1r\rangle-|r1\rangle)/\sqrt{2}$ are excluded as they do not play a role in the driven dynamics.

During the echoing RAP processes, the input $|11\rangle$ state evolves adiabatically along the 
instantaneous eigenstate [blue line with an arrow in Fig.~\ref{fig:CZgate}(b) to emphasize the evolution direction]. 
After the first RAP, the system reaches the $(|1r\rangle + |r1\rangle)/\sqrt{2}$ state indicating the coherent population transition,
as revealed by the population dynamics (solid lines) in left pannel of Fig.~\ref{fig:CZgate}(c). 
Succeeding the second RAP, the dynamical state returns back to $|11\rangle$ adiabatically 
[c.f.  Fig.~\ref{fig:CZgate}(b) and solid lines in right pannel of Fig.~\ref{fig:CZgate}(c)], 
with extra dynamical and geometric phases left. 
The dynamical phases, $\phi_d^{k=1,2}=\int E_k(t) dt$, 
accumulated during the first RAP are depicted by the red shaded region in Fig.~\ref{fig:CZgate}(b), 
and during the second by the blue shaded region, 
which can exactly cancel out each other when Rydberg blockade is perfectly satisfied. 
The Rydberg blockade plays a crucial role in maintaining the symmetry of the adiabatic spectrum, as shown in Fig.~\ref{fig:CZgate}(b). 
It ensures a nearly two-level coupling process, similar to a Landau-Zener process, 
by decoupling the $|rr\rangle$ state from $|1r\rangle+|r1\rangle$. 
When the blockade is reduced, the symmetry of the adiabatic spectrum becomes distorted. 
This distortion leads to imperfect cancellation of the dynamical phases, 
ultimately reducing the fidelity of the gate.
Moreover, due to the cyclic evolution after two RAP processes, an overall geometric phase is acquired~\cite{Chen20high,Bosch2023Shortcut,Yu19OE}, 
which is half of the solid angle enclosed by the evolution path, 
and is $\pi$ here for the target CZ gate~\cite{Beterov_2020,Pelegri_2022}.
The dynamics of input $|11\rangle$ state during the echoing RAP protocol is dipicted in the inset of Fig.~\ref{fig:CZgate}(c).
The state evolution covers a full cycle with a solid angle of $\pi$, 
demonstrating the echoing dynamics of coherent population inversions with central symmetry around the sphere center.
Similarly, if the input state is $|10\rangle$ ($|01\rangle$), 
the state will evolve adiabatically along the red line in Fig.~\ref{fig:CZgate}(b)
to the $|r0\rangle$ ($|0r\rangle$) state, as the $|0\rangle$ state does not participate in the evolution,
and then return to the $|10\rangle$ ($|01\rangle$) state after the second RAP pulse. 
The dynamics of population on single-atom $|r\rangle$ and $|1\rangle$ states are depicted by dashed lines in Fig.~\ref{fig:CZgate}(c).
Like the case of $|11\rangle$ state, the dynamical phases accumulated after the two RAP pulses cancel out with each other, 
leaving only the geometric phase of $\pi$. 
Unlike entering the $|11\rangle$ state, this single-atom process is independent of the interaction strength, 
yielding the same result under different coupling conditions.
The key point is that the dynamical phases accumulated are different for all computational bases but cancel out during the two RAP pulses due to the symmetry of the eigenenergy spectrum in Fig.~\ref{fig:CZgate}(b).
The above echoing RAP scheme accomplish the CZ gate: $\mathcal{U}_{\rm CZ}=\text{diag}(1, -1, -1, -1)$, 
with computational basis $\{|00\rangle, |01\rangle, |10\rangle, |11\rangle\}$.

The design of the three-qubit CCZ gate follows a similar approach to that of the CZ gate, 
with $\Omega(t)$ and $\Delta(t)$ fine-tuned and indicated by dashed lines in Fig.~\ref{fig:CZgate}(a).
Three qubits are arranged in an equilateral triangle to ensure symmetry within the Rydberg blockade regime between each pair of atoms~\cite{e24101371,Evered2023}. 
Additionally, a strong Rydberg interaction between adjacent atoms is necessary to accomplish the cancellation of the dynamical phase.
The logic transformation of input states can be described as follows:
(i) For input state $|111\rangle$, one has
$|111\rangle \xrightarrow{\rm RAP} (|r11\rangle+|1r1\rangle+|11r\rangle)/\sqrt{3} \xrightarrow{\rm RAP} -|111\rangle.$
(ii) The states $|011\rangle$, $|101\rangle$, and $|110\rangle$ undergo the same evolution process as the previously 
described double-atom case of $|11\rangle$, as only two of the three atoms are populated in RAP pulse-coupled states.
(iii) The states $|001\rangle$, $|010\rangle$, and $|100\rangle$ with only one atom in state $|1\rangle$ evolve similarly to the single-atom case of $|1\rangle$. 
(iv) The state $|000\rangle$ is dark to the echoing RAP pulses. 
Hence, following two RAP processes, all states except $|000\rangle$ effectively cancel the dynamical phase 
and accumulate a geometric phase of $\pi$.
Consequently, the matrix representation of the CCZ gate is given by 
$\mathcal{U}_{\rm CCZ} = \text{diag}(1,-1,-1,-1,-1,-1,-1,-1)$ with respect to the computational 
basis $\{|000\rangle$, $|001\rangle$, $|010\rangle$, $|011\rangle$, $|100\rangle$, $|101\rangle$, $|110\rangle$, $|111\rangle\}$.

\section{Gate fidelity and robustness}\label{sec:fidelity}

\subsection{Realistic atom parameter implementation}\label{subsec:akali}
We now ultilize realistic parameters for experiments involving alkali-metal atoms to assess 
the efficacy of our echoing protocol.
In experiments carried out in a 1D tweezer array of neutral atoms, 
the Rydberg interaction strength can be on the order of  $\sim$GHz to $\sim$THz for akali-metal atoms such as
$^{133}$Cs and $^{87}$Rb~\cite{Graham19Rydberg,saffman2020symmetric, levine2019parallel,McDonnel22Demonstration,Graham23midcircuit}.
Futhermore, to consider spontaneous emissions and dephasing of the Rydberg state $|r\rangle$,
we simulate the system dynamics with optimized RAP profiles using the Lindblad master equation:
\begin{equation}\label{eq:mastereq}
  \partial_t\hat{\rho}(t)=-i[\hat{H}(t), \hat{\rho}]+\sum_k\mathcal{D}_k[\hat{\rho}],
\end{equation}
where the dissipators are given by
\begin{equation*}
  \mathcal{D}_k[\hat\rho]=\hat{L}_k\hat{\rho}\hat{L}_k^\dagger-\frac{1}{2}(\hat{L}_k^\dagger \hat{L}_k\hat{\rho}+\hat{\rho}\hat{L}_k^\dagger \hat{L}_k),
\end{equation*}
and $\{L_k\}$ are jump operators describing the dissipative quantum channels.
For the $|r\rangle=|107P_{3/2}; m_J = 3/2\rangle$ Rydberg state of $^{133}$Cs,
we use: $\hat{L}_1=\sqrt{\gamma_{r}/16}|0\rangle\langle r|$, 
$\hat{L}_2=\sqrt{\gamma_{r}/16}|1\rangle\langle r|$
for spontaneous emission to ground state manifold,
and $\hat{L}_3=\sqrt{7\gamma_{r}/8}|r\rangle\langle r|$ for dephasing channel,
and the total dissipation rate is $\gamma_r=1/(540 \,\mu s)$~\cite{saffman2020symmetric}. 
We optimize variational parameters in Eq.~(\ref{eq:omega_delta}), which are then used to simulate Eq.~(\ref{eq:mastereq}) and evaluate its performance with dissipations. 
The optimized parameters are $\Delta_0=0.55 \Omega_{\rm max}$, $(\tau_R, \tau_D)=(0.564,0.846)$, $\tau_{\rm gate}=\Omega_{\rm max} t=9.4$, $V_0=127\Omega_{\rm max}$ for the CZ gate, 
and $\Delta_0=0.70 \Omega_{\rm max}$, $(\tau_R, \tau_D)=(0.72,1.08)$, $\tau_{\rm gate}=12.0$, $V_0=100\Omega_{\rm max}$ for the CCZ gate.

\begin{figure}
  \centering
  \includegraphics[width=1\columnwidth]{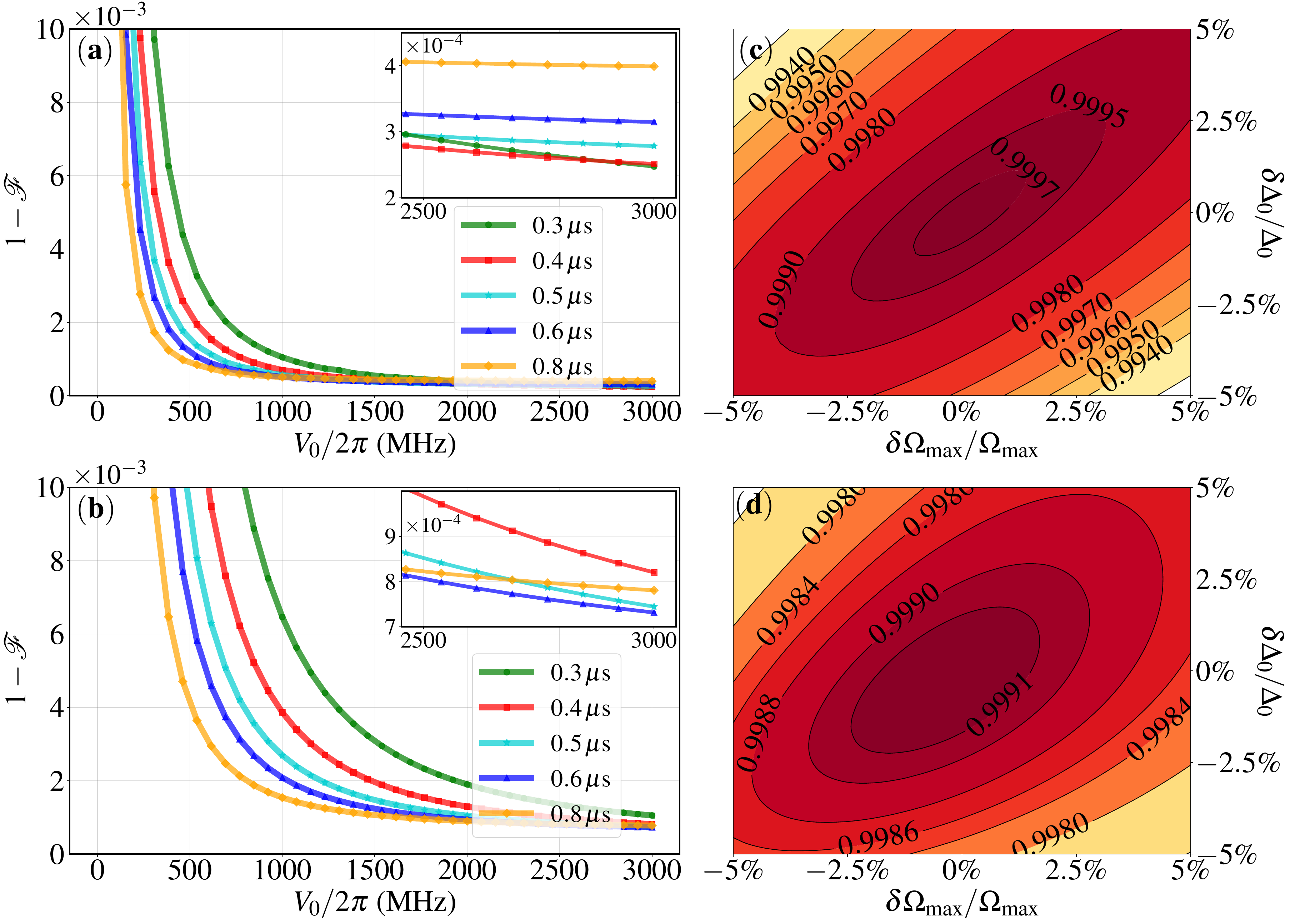}
  \caption{(a) and (b): Variation of the infidelity with $V_0$ at different total times 
  for the CZ and CCZ gates, respectively. 
  The inset zooms in on $V_0/2\pi \in [2500, 3000]$ MHz. 
  In (a), the red stars represent the data from Ref.~\cite{saffman2020symmetric} with a total time of 0.54 $\mu s$, 
  using the same atomic Rydberg state and decay rate. 
  (c) and (d): Contour plots of the fidelity for the CZ and CCZ gates against parameter fluctuations, 
  with $\Omega_{\rm max} = (2\pi)\,23.5$ MHz for CZ gate and $\Omega_{\rm max} = (2\pi)\,30$ MHz for CCZ gate, resulting in both total gate time of 0.4\,$\mu$s. 
  }
\label{fig:CZ_V0}
\end{figure}

Figure~\ref{fig:CZ_V0}(a) assesses the fidelity of the engineered CZ gate across varying Rydberg 
interaction strengths $V_0$ for different gate times $t_{\rm gate}$, 
employing parameters specific to $^{133}$Cs atoms.
For  qubits encoded into the clock states of $^{133}$Cs, 
we predict that our protocol realize a CZ gate with duration of less than $1\,\mu$s,
accompanied by an infidelity on the order of $10^{-3}$.
From the comparison with the red stars representing the scheme in Ref.~\cite{saffman2020symmetric} in Fig.~\ref{fig:CZ_V0}(a), 
it can be observed that our scheme outperforms the original protocol for both low and high interaction strengths.
As revealed in Figs.~\ref{fig:CZ_V0}(a) and (b), the controlled phase error primarily arises from 
undesired cancellation of dynamical phases, 
which gradually decreases as the Rydberg interaction strength $V_0$ increases.
All schemes can achieve a fidelity of approximately around 0.9997 when the interaction strength reaches $V_0 = (2\pi)\,3$ GHz
(near the maximum possible for the Cs $|107P_{3/2}; m_J = 3/2\rangle$ state~\cite{saffman2020symmetric}).
In the schemes presented here for CZ gate, we optimize the gate fidelity for experimentally achievable parameters, 
ensuring that the Rabi frequency is below $(2\pi)\,35$ MHz and the maximum laser detuning is below $(2\pi)\,40$ MHz. 
In practice, one can achieve the desired fidelity by balancing various experimental conditions, such as interaction strength, 
total gate time, and pulse intensity, to suit specific system requirements.
In near-term experimental settings, utilizing high Rydberg states (e.g., $n=107$) 
and achieving high Rabi frequencies may present challenges, 
leading to the use of more modest parameters. 
To provide a reference for fidelity under these conservative conditions, 
we have calculated the fidelity using a Rabi frequency of $(2\pi)\,5$ MHz, 
a Rydberg lifetime of 130 $\mu s$, and a Rydberg-Rydberg interaction strength of only $(2\pi)\,400$ MHz. 
Under these parameters, we obtain a fidelity of 0.9963, which can serve as a benchmark for actual experiments.
Figure~\ref{fig:CZ_V0}(b) depicts the fidelity of the CCZ gate for various time schemes as a function of $V_0$. 
It is evident that at the maximum interaction strength of $V_0 = (2\pi) 3$ GHz for Cs atoms \cite{saffman2020symmetric}, 
all CCZ gate scenarios achieve a fidelity exceeding 0.999. 
Here, the Rabi frequency is maintained below $(2\pi)$40 MHz, and the maximum detuning is limited to $(2\pi)$\,50 MHz.

The tweezer array with alkaline earth atoms in metastable excited states has unlocked new capabilities, 
such as tweezer-based atomic clocks and long-lived nuclear spin qubits~\cite{Chen22Analyzing,Huie23repeti, Ma2023,Ma22PRX,Jenkins22PRX,Norcia23midcircuit,Lis23Midcircuit}. 
We further analyze our scheme with $^{171}$Yb qubits using single-photon excitation to the $|75\,{^3}S_{1}; F = 3/2\rangle$ Rydberg state
(lifetime $100\,\mu\text{s}$). 
The specific numerical values considered here are $V_0=(2\pi)\,6.85$ GHz for an atom distance of 3\,$\mu$m~\cite{Jandura23PRXQ}. 
With $\Omega_{\rm max} = (2\pi)\,23.5$ MHz for CZ gate and $\Omega_{\rm max} = (2\pi)\,30$ MHz for CCZ gate, resulting in both total gate time of 0.4 $\mu$s, 
we predict a CZ fidelity of 0.9984 and CCZ fidelity of 0.9979 using our echoing RAP protocol. 
These results demonstrate that our scheme maintains high fidelity and excellent adaptability for different atoms.
This capability may facilitate the generation of entanglement for alkaline earth atoms in clock transitions, 
thereby enhancing metrology and other related applications~\cite{Schine2022Long, cao2024multiqubit}.

\subsection{Robustness to experimental imperfections}\label{subsec:roubust}
We now proceed to explore how our protocol responds to various imperfections. 
One particular aspect of interest is the effect of laser intensity variation on gate fidelity.
Experimentally, such variations can arise from calibration errors or slow drifts in the laser system during the gate execution period~\cite{Sensitivity2023Jiang}. 
In the context of two-photon excitation schemes, fluctuations in laser intensity can also introduce additional errors in detuning~\cite{Fromonteil23Protocols}.
Another factor contributing to gate errors in experiments involving 
Rydberg atom arrays is the thermal motion of atoms in the traps~\cite{Graham19Rydberg,Graham23midcircuit}. 
This motion induces a Doppler shift, resulting in shot-to-shot fluctuations in the effective laser detuning.
To analyze these situations, we consider $\delta\Omega_{\rm max}/\Omega_{\rm max}$ and $\delta\Delta_0/\Delta_{0}$ 
to assess the robustness of our echoing protocol, 
where $\delta\Omega_{\rm max}$ and $\delta\Delta_0$ represent the deviations of 
the Rabi frequency and laser detuning from their optimal values, respectively.

As an illustrative example, the robustness of our scheme for a 0.4 $\mu$s gate duration with $V_0/2\pi = 3$ GHz is depicted in Figs.~\ref{fig:CZ_V0}(c) and (d).
Figure \ref{fig:CZ_V0}(c) illustrates the fidelity variation of the CZ gate with changing parameters. 
It shows that even with approximately a $\pm1$\% variation in detuning and a $\pm3$\% variation in Rabi frequency, 
the fidelity remains at 0.9990. 
This level of tolerance allows for a Doppler detuning variation around $\pm300$\,kHz, 
which exceeds the typical $\pm80$\,kHz Doppler-shift variation observed for a Cs atom cooled to 10\,$\mu$K 
and excited to a Rydberg state via a one-photon transition~\cite{saffman2020symmetric,Fromonteil23Protocols}.
The Doppler shift manifests as a frequency offset error for $\Delta(\tau)$. 
We checked that with a $\pm 100$ kHz detuning offset error, 
the CZ gate fidelity remains above 0.9965 and the CCZ gate fidelity remains above 0.9968,
indicating that the three-qubit gate is less susceptible to errors than the two-qubit gate~\cite{Jin24Geometric}.
For the CCZ gate in Fig.~\ref{fig:CZ_V0}(d), 
the fidelity drops to 0.997 with a $\pm5\%$ variation in both Rabi frequency and detuning, maintaining high fidelity.
Additionally, errors in the system evolution time can affect the fidelity of our scheme. 
The fidelity remains above 0.999 for a time variation of up to $\pm 5\%$ (or $\pm 20$ ns) 
around the optimized evolution time, demonstrating the scheme's robustness. 
In a practical experimental system, the accurate control of the evolution time in a gate operation is typically realized by an arbitrary waveform generator (AWG) with sampling rate no less than 1GS/s, 
which is capable to achieve temporal accuracy within a few tens of nanoseconds, 
making the fidelity reduction due to timing errors negligible and acceptable.

The echoing RAP pulse has the potential to directly realize $\text{C}^k\text{Z}$ gates with more control qubits. 
When positioning four atoms at the vertices of an equilateral pyramid~\cite{Barredo18Synthetic} with blockade strength $V_0=(2\pi)\,3$\,GHz, 
and utilizing optimized parameters for Cs atoms without further optimization for the CCZ gate in Fig.~\ref{fig:CZ_V0}(d), 
we predict a CCCZ gate fidelity of 0.997. 
Alternatively, placing the four atoms at the vertices of a square in a plane yields the fidelity of 0.996.
Thus far, successful construction of multi-qubit $\text{C}^k\text{Z}$ gates has been achieved. 
These gates, combined with single-qubit operations, form $\text{C}^k\text{NOT}$ gates, 
enabling the generation of quantum entangled states and other state preparation processes, 
and serve as fundamental units in a universal quantum circuit~\cite{Lu2019, Most07Formation, Guan14Entangling}.

\section{Conclusion and Discussion}\label{sec:conclusion}
In summary, we propose to use echoing RAP pulses with global laser dressing to achieve high-fidelity 
controlled-Z gates with Rydberg atoms. 
A controlled phase gate can also be implemented by introducing a change in the laser phase~\cite{Fromonteil23Protocols}.
The results demonstrate robust performance and resilience to parameter variations, 
showcasing potential for constructing multi-qubit gates efficiently. 
This advancement holds promise for enhancing quantum information processing capabilities 
and improving the efficiency of quantum computation.
The uniform and continuous application of pulses across all atoms in our scheme may eliminate the need for rapid laser switching between different spatial locations.
Our scheme promises CZ gates with high fidelity, 
greater resilience to parameter variations, 
and direct extension to C$^k$Z gates without further optimization, all within a shorter gate time.

Notably, the convergence of programmable Rydberg-atom arrays with optical atomic clocks presents a unique opportunity. 
Our scheme can also be implemented in the optical-metastable states of alkaline-earth atoms, 
such as Ytterbium clock states~\cite{Chen22Analyzing}, 
or a Strontium fine-structure qubit via a single-photon process~\cite{Unnikrishnan24coherent,pucher24fine}, 
promising advancements in entanglement-enhanced metrology~\cite{Schine2022Long, Shaw24multi,finkelstein2024universal,cao2024multiqubit}.

\section*{Acknowledgements}
We thank the anonymous referees for their helpful feedback, 
which improved our presentation of results.
The authors thank Qi Liu for useful discussions and comments on the manuscript. 
This work was supported by Innovation Program for Quantum Science and Technology (Grant No.\,2021ZD0302100),
National Natural Science Foundation of China (Grant No.\,12304543 and No.\,92265205), and
Open Research Fund Program of the State Key Laboratory of Low-Dimensional Quantum Physics (KF202111).
M.X. was also supported by Innovative and Entrepreneurial Talents Project (No.\,JSSCBS20220234).
S.X. was supported by Innovation Training Program of NUAA (No.\,20241007700100Z and No.\,2024CX021030).

\bibliography{mybib}

\end{document}